\documentclass[aps,prx,twocolumn,preprintnumbers,amsmath,amssymb,superscriptaddress]{revtex4-2}
\usepackage{graphicx}
\usepackage{amsmath}
\usepackage{color}
\usepackage[sort&compress]{natbib}
\begin{document}

\title{Ground state and magnetic transitions of the orthorhombic antiferromagnet CaCo$_2$TeO$_6$}

\author{Xing Huang}
\affiliation{Center for Neutron Science and Technology, Guangdong Provincial Key Laboratory of Magnetoelectric Physics and Devices, School of Physics, Sun Yat-sen University, Guangzhou, Guangdong 510275, China}
\author{Peiyue Ma}
\affiliation{Center for Neutron Science and Technology, Guangdong Provincial Key Laboratory of Magnetoelectric Physics and Devices, School of Physics, Sun Yat-sen University, Guangzhou, Guangdong 510275, China}
\author{Mengwu Huo}
\affiliation{Center for Neutron Science and Technology, Guangdong Provincial Key Laboratory of Magnetoelectric Physics and Devices, School of Physics, Sun Yat-sen University, Guangzhou, Guangdong 510275, China}
\author{Chaoxin Huang}
\affiliation{Center for Neutron Science and Technology, Guangdong Provincial Key Laboratory of Magnetoelectric Physics and Devices, School of Physics, Sun Yat-sen University, Guangzhou, Guangdong 510275, China}
\author{Hualei Sun}
\affiliation{School of Science, Sun Yat-Sen University, Shenzhen, Guangdong 518107, China}
\author{Xiang Chen}
\affiliation{Center for Neutron Science and Technology, Guangdong Provincial Key Laboratory of Magnetoelectric Physics and Devices, School of Physics, Sun Yat-sen University, Guangzhou, Guangdong 510275, China}
\author{Sihao Deng}
\affiliation{Spallation Neutron Source Science Center, Dongguan, Guangdong 523803, China}
\author{Lunhua He}
\affiliation{Spallation Neutron Source Science Center, Dongguan, Guangdong 523803, China}
\author{Tao Xie}
\affiliation{Center for Neutron Science and Technology, Guangdong Provincial Key Laboratory of Magnetoelectric Physics and Devices, School of Physics, Sun Yat-sen University, Guangzhou, Guangdong 510275, China}
\author{Zhangzhen He}
\email{hezz@fjirsm.ac.cn}
\affiliation{State Key Laboratory of Structural Chemistry, Fujian Institute of Research on the Structure of Matter, Chinese Academy of Sciences, Fuzhou, Fujian 350002, China}
\author{Meng Wang}
\email{wangmeng5@mail.sysu.edu.cn}
\affiliation{Center for Neutron Science and Technology, Guangdong Provincial Key Laboratory of Magnetoelectric Physics and Devices, School of Physics, Sun Yat-sen University, Guangzhou, Guangdong 510275, China}

\begin{abstract}
We report the systematic synthesis, crystal structure, magnetization, and powder neutron diffraction of single crystalline and polycrystalline CaCo$_2$TeO$_6$ samples. CaCo$_2$TeO$_6$ crystallizes in an orthorhombic structure with $Pnma$ space group, featuring chains of edge-shared CoO$_6$ octahedra arranged in a honeycomb pattern. Two antiferromagnetic transitions are observed at $T$$_{N1}$ = 14.4 K and $T$$_{N2}$ = 16.2 K, corresponding to two long-range magnetic orders with propagation vectors of $\bf{k}$$_1$ = (0, 0, 0) and $\bf{k}$$_2$ = (0.125, 0, 0.25), respectively. The ground state is determined as a canted up-up-down-down zigzag spin configuration along the $c$ axis, wherein the magnetic moments of Co1 and Co2 ions are 3.4(1) and 2.1(1)$\mu$$_B$, respectively. Successive spin-flop transitions appear with the increasing magnetic field applied along the easy axis ($c$ axis), accompanied by depression of the antiferromagnetic orders and enhancement of residual magnetic entropy. The field-induced spin-disordered state suggests that CaCo$_2$TeO$_6$ may be an ideal candidate for studying frustrated magnetism.
\end{abstract}

\maketitle

\section{INTRODUCTION}
The investigation of quantum magnetism in low-dimensional magnets has become a central theme in modern condensed matter research due to the tunable frustration and enhanced quantum fluctuations. Honeycomb lattice with structural simplicity (coordination number $z$ = 3) stands out as a paradigmatic platform. The pristine honeycomb lattice with nearest-neighbor Heisenberg interaction ($J$$_1$) supports classical N\'{e}el antiferromagnetism without inherent frustration, its deliberate modification through additional exchange pathways opens avenues for ground state control\cite{PhysRevB.86.144404, PhysRevLett.110.127203}. Introducing competing interactions ($J$$_2$ and $J$$_3$) transforms this system into a frustrated $J$$_1$-$J$$_2$-$J$$_3$ network, which can give rise to various ground states, such as spiral order\cite{PhysRevB.81.214419}, plaquette valence bond state\cite{PhysRevB.96.104401}, and spin liquid\cite{PhysRevLett.117.167201}.

Research on honeycomb magnets has been prompted by Kitaev in an alternative avenue, who gave a precisely solvable Kitaev model with bond-dependent Ising interactions on a honeycomb lattice\cite{kitaev2006anyons, jackeli2009mott, savary2017quantum, knolle2019field}. In honeycomb materials with effective $S$ = 1/2 and spin-orbit coupling (SOC)\cite{witczak2014correlated, cao2018challenge}, the Kitaev model yields a gapped quantum spin liquid (QSL) ground state, which hosts fractionalized Majorana fermion excitations intertwined with a $Z$$_2$ gauge fluxes\cite{banerjee2017neutron, takagi2019concept}. The 4$d$$^5$ $\alpha$-RuCl$_3$\cite{banerjee2017neutron, yokoi2021half, banerjee2018excitations, kasahara2018unusual, banerjee2016proximate, Baek2017, Sears2017, Wolter2017, Ran2017, Zhao_2022, Winter2018Probing} and 5$d$$^5$ $A$$_2$IrO$_3$ ($A$ = Na, Li)\cite{Takayama2015, Mehlawat2017} families with strong SOC are the well-known QSL candidates, exhibiting many intriguing physical properties such as fractionalized excitations and anomalous thermal Hall effect.

Recently, the interests extended to the 3$d$$^7$ Co$^{2+}$ compounds with a $t$$_{2g}^5$$e$$_{g}^2$ configuration ($S$ = 3/2, $L$ = 1). The SOC in Co$^{2+}$ ions stabilizes a pseudospin-1/2 ground state and generates pronounced magnetic anisotropy in exchange interactions, establishing a promising platform for realizing the Kitaev QSL\cite{liu2018pseudospin, sano2018kitaev, liu2020kitaev}. The honeycomb cobaltates $A$$_2$Co$_2$TeO$_6$ ($A$ = Na, K)\cite{yao2020ferrimagnetism, songvilay2020kitaev, lin2021field, chen2021spin, samarakoon2021static, yao2022excitations, yao2023magnetic, xiang2023disorder, francini2024spin, krueger2023triple, xu2024k2co2teo6}, BaCo$_2$(AsO$_4$)$_2$\cite{zhong2020weak, regnault2018polarized, Thomas2023, zhang2023magnetic, shi2021magnetic}, and Na$_3$Co$_2$SbO$_6$ \cite{kim2022antiferromagnetic, li2022giant, gu2024in,songvilay2020kitaev, samarakoon2021static} all exhibit antiferromagnetic ground states. Intriguingly, under a properly aligned magnetic field, they all transition to a spin-disordered state, the so-called QSL-like disordered state or quantum paramagnetic state, a candidate state of the QSL. CaCo$_2$TeO$_6$ polycrystals obtained via the aliovalent topochemical ion exchange method show a two-dimensional honeycomb structure with a weak trigonal distortion. They exhibit a spin-disordered state under a relatively low magnetic field of  $\sim$4 T, suggesting a Kitaev QSL candidate\cite{PhysRevB.110.184411}. These discoveries motivated us to investigate its magnetic properties deeply through anisotropic measurements using single crystals.

We have successfully grown single crystals of CaCo$_2$TeO$_6$ by flux method. The single crystals show a different structure due to the trigonal distortion, which consists of chains of edge-shared CoO$_6$ octahedra in a honeycomblike arrangement. Two antiferromagnetic (AFM) transitions are observed at 14.4 and 16.2 K, respectively, and the Curie-Weiss fittings suggest strong AFM couplings. Neutron powder diffraction (NPD) reveals a canted up-up-down-down ($\uparrow$$\uparrow$$\downarrow$$\downarrow$) zigzag spin configuration along the $c$ axis. The easy-axis magnetic field can drive this system to undergo successive spin-flop transitions. Unexpectedly, a field-induced spin-disordered state is also observed in the single crystal samples.

\section{EXPERIMENTAL DETAILS}

Single crystals of CaCo$_2$TeO$_6$ were grown by a flux method. The starting materials CaCO$_3$ (99.99\%), CoC$_2$O$_4$$\cdot$2H$_2$O (99.9\%), and TeO$_2$ (99.99\%) were mixed thoroughly in a molar ratio 1 : 2 : 2 in an agate mortar and filled into a platinum crucible. The crucible was heated to 1080 $^{\circ}$C, kept for 10 h, cooled to 1000 $^{\circ}$C at 3.3 $^{\circ}$C/h, then cooled down to room temperature at 40 $^{\circ}$C/h. The as-grown purple crystals were mechanically separated. Polycrystalline samples of CaCo$_2$TeO$_6$ were prepared using the conventional solid-state reaction. The starting materials CaCO$_3$ (99.99\%), Co$_3$O$_4$(99.99\%), and TeO$_2$ (99.99\%) were mixed thoroughly in the stoichiometric ratio in an agate mortar and then pressed into a pellet. The pellet was put into an alumina crucible covered with a lid. The pellet was first heated at 800 $^{\circ}$C and 900 $^{\circ}$C for 5 h, respectively, and then heated at 1000 $^{\circ}$C for 3 d with several intimidate grindings.

Single-crystal x-ray diffraction (XRD) was measured on a SuperNova (Rigaku) single-crystal x-ray diffractometer. The crystal structure was solved utilizing the SHELXTL software package\cite{Sheldrick1998}, and the refined structural parameters are summarized in Table I. NPD was conducted on the general purpose powder diffractometer (GPPD) installed in the China Spallation Neutron Source (CSNS). Polycrystalline samples of CaCo$_2$TeO$_6$ were filled in cylindrical Titanium Zirconium alloy can and measured at 6, 8, 10, 11, 13, 15, 16, 18, and 230 K. Rietveld refinements and magnetic structure analysis for the NPD patterns were performed by the GSAS-II and K-SUBGROUPSMAG programs\cite{Toby2013, PerezMato2015}. The elemental analysis was performed using an energy dispersive x-ray spectroscopy (EDX) (EVO, Zeiss). The magnetic susceptibility and heat capacity measurements were carried out in a physical property measurement system (PPMS, Quantum Design).

\begin{figure}[t]
\includegraphics[scale = 1]{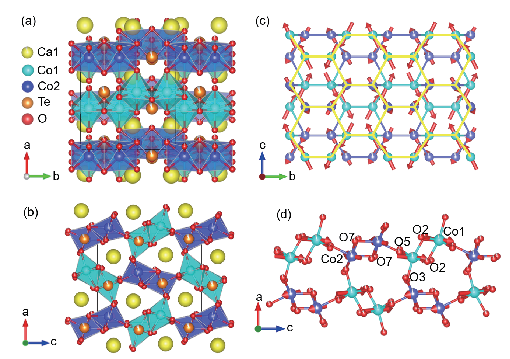}
\caption{ Crystal structure of the orthorhombic CaCo$_2$TeO$_6$ viewed along (a) the $c$ axis and (b) the $b$ axis. (c) Sketch of the magnetic structure in the ${bc}$ plane refined by NPD data of 6 K. The Co$^{2+}$-based honeycomb layered structure is presented with yellow hexagons. (d) The zigzag chains of Co$^{2+}$ ions along the $c$ axis.}
\label{fig1}
\end{figure}

\begin{table}[t]
\caption{Atoms, Wyckoff sites, fractional atomic coordinates, and occupations of CaCo$_2$TeO$_6$ from the NPD at 230 K. The refined lattice constants are $a$ = 9.2662(4), $b$ = 8.9771(3), and $c$ = 10.8809(4) {\AA}, and the space group is orthorhombic $Pnma$. The fitting quality is given by a $R$$_{wp}$ = 1.85\%, and a GOF = 1.43.}
\begin{tabular}{ccccccc}
\hline \hline
Atoms  &Wyck.  & $x$       & $y$       & $z$             & Occ. \\ \hline
Te1    &4c     & 0.5842(3) & 0.25    & 0.8957(3)       & 0.911(9) \\
Te2    &4c     & 0.4392(3) & 0.75    & 0.6233(3)      & 0.927(9)\\
Co1    &8d     & 0.5837(4) & 0.5903(5) & 0.8926(3)          & 0.98(1)  \\
Co2    &8d     & 0.5679(4) & 0.9094(5) & 0.3925(4)          & 0.95(1) \\
Ca1    &8d     & 0.7585(2) & 1.0483(2) & 0.6553(3)         & 1  \\
O1     &4c     & 0.4023(3) & 0.75    & 0.4454(3)          & 0.93(1)    \\
O2     &8d     & 0.6213(2) & 0.4085(3) & 1.0085(2)        & 0.941(8) \\
O3     &8d     & 0.3009(7) & 0.5910(2) & 0.6556(2)         & 0.973(7) \\
O4     &4c     & 0.3889(3) & 0.25   & 0.9578(3)         & 0.96(1)   \\
O5     &8d     & 0.5322(2) & 0.0929(2) & 0.7810(2)         & 0.966(8)     \\
O6     &4c     & 0.7784(3) & 0.25    & 0.8287(3)        & 0.957(7)   \\
O7     &8d     & 0.5736(2) & 0.9076(3) & 0.5805(2)     & 0.957(7)     \\
O8     &4c     & 0.5044(3) & 0.75    & 0.7864(3)       & 0.97(8)     \\ \hline \hline
\end{tabular}
\label{table:t1}
\end{table}

\begin{figure}[t]
\includegraphics[scale = 1]{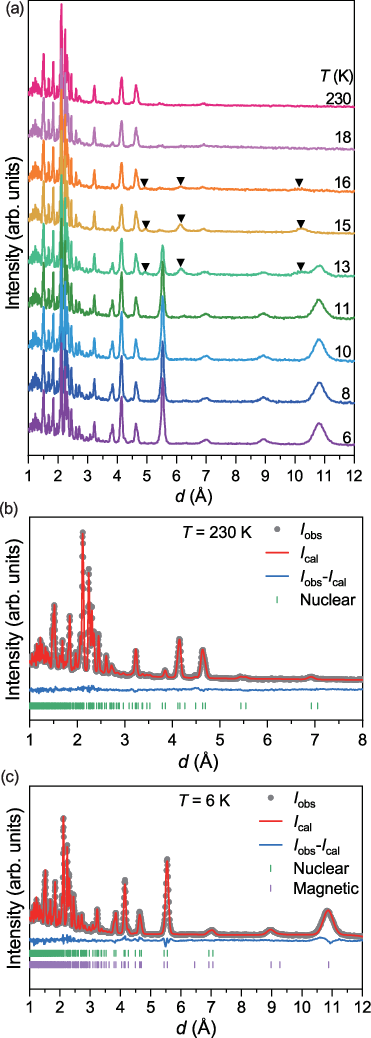}
\caption{(a) NPD patterns of CaCo$_2$TeO$_6$ at various temperatures. The black triangles mark the magnetic diffraction peaks of the intermediate phase ($T$$_{N1}$ $<$ $T$ $<$ $T$$_{N2}$). Rietveld refinement results of the NPD patterns at (b) 230 K and (c) 6 K.}
\label{fig2}
\end{figure}

\section{RESULTS AND DISCUSSIONS}

Figure \ref{fig1} shows the structure of CaCo$_2$TeO$_6$ with an orthorhombic $Pnma$ space group. The lattice parameters refined from the NPD in Fig. \ref{fig2} measured at $T$ = 230 K are $a$ = 9.2662(4), $b$ = 8.9771(3), and $c$ = 10.8809(4) {\AA}. The refined structural parameters are summarized in Table 1. The refined composition CaCo$_{1.94}$Te$_{0.92}$O$_{5.76}$ is close to the EDX result (Ca: Co: Te = 1: 2: 0.92) of a single crystal sample, consistent with the stoichiometry of CaCo$_2$TeO$_6$. There are two crystallographic sites of Co$^{2+}$ ions in a unit cell, forming distorted CoO$_6$ octahedra. As illustrated in Figs. \ref{fig1}(a) and \ref{fig1}(b), the CoO$_6$ octahedra form a double chain structure running along the $b$ axis, and the chains link to each other by the corner-sharing oxygen in a honeycomblike arrangement to construct a three-dimensional structural framework. Each Co$^{2+}$ ion is surrounded by four neighboring Co$^{2+}$ ions. The distances between adjacent Co$^{2+}$ ions within the chains range from 2.862(9) to 3.240(8) {\AA}, and the shortest interchain distances are 3.614(6) and 3.773(6) {\AA} along the $a$ and $c$ axis, respectively. The Ca$^{2+}$ cations are in the interchain channels. This crystal structure is also confirmed by a single crystal XRD experiment (see the Supplemental Material)\cite{x1}. This structure differs from the recently reported polycrystalline phase obtained by ion-exchange reaction, which is crystallized in the rhombohedral $R$$\bar{3}$ with Co$^{2+}$-two-dimensional honeycomb layers well separated by interlayer Ca$^{2+}$ cations\cite{PhysRevB.110.184411}.

To further investigate the magnetic ground state of the orthorhombic CaCo$_2$TeO$_6$, NPD is performed in the temperature range from 6 to 230 K. Figure \ref{fig2} shows the NPD patterns and their refinements. Compared with the NPD patterns at 230 K, three extra peaks appear at 16 K, which feature relatively large $d$ values and can be assigned as magnetic Bragg peaks. These magnetic peaks can be indexed with a propagation vector $\bf{k}$$_2$ = (0.125, 0, 0.25) and show a maximum intensity at 15 K. As further cooling, these magnetic peaks are gradually suppressed, and a series of new extra magnetic peaks can be observed below 13 K, which can be indexed by a propagation vector of $\bf{k}$$_1$ = (0, 0, 0). The new magnetic reflections with $\bf{k}$$_1$ and $\bf{k}$$_2$ evidence the formation of long-range magnetic orders. Figure \ref{fig2}(c) shows the NPD pattern and its refinement at 6 K. The refined magnetic space group is $Pnma$, and the magnetic moments along the $a$, $b$, and $c$ directions are $m$$_a$ =  0.23(8), $m$$_b$ = -1.53(4), and $m$$_c$ =  3.03(2)$\mu$$_B$ for Co1, and $m$$_a$ =  0.44(8), $m$$_b$ = 1.08(3), and $m$$_c$ =  1.84(1)$\mu$$_B$ for Co2. The total moments are 3.4(1) and 2.1(1)$\mu$$_B$ for Co1 and Co2, respectively. Figure \ref{fig1}(c) shows the corresponding magnetic structure. The Co$^{2+}$ ions also form honeycomb layers stacking in an $ABAB$ manner along the $a$ axis direction, and spins are arranged in the ${bc}$ planes. As shown in Fig. \ref{fig1}(d), the spins along the $c$ axis form alternating FM and AFM zigzag chains, and the adjacent AFM spins are oriented by a 4.6$^\circ$ angle, which can be viewed as a canted $\uparrow$$\uparrow$$\downarrow$$\downarrow$ spin configuration. Compared with the NPD results below $T$$_{N1}$, the intermediate phase ($T$$_{N1}$ $<$ $T$ $<$ $T$$_{N2}$) has rather weaker magnetic reflection peaks and larger magnetic period, making it difficult to determine the magnetic structure.

\begin{figure*}[t]
\centering
\includegraphics[scale = 0.85]{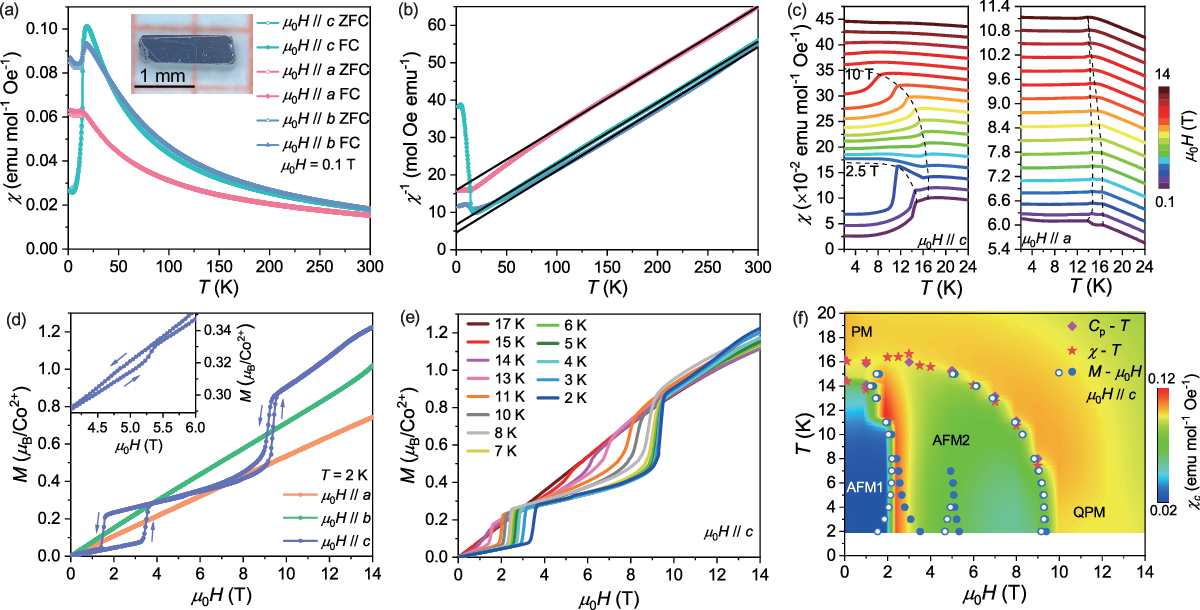}
\caption{(a) Temperature dependence of zero-field-cooled (ZFC) and field-cooled (FC) magnetic susceptibilities measured with a 0.1 T magnetic field along the $a$, $b$, and $c$ axis. The inset shows a single crystal of the orthorhombic CaCo$_2$TeO$_6$. (b) Curie-Weiss fit of the inverse susceptibility data from 100 to 300 K. (c) Temperature-dependent magnetic susceptibility under different magnetic fields for $H$ $\|$ $c$ (left) and $H$ $\|$ $a$ (right). The data are shifted upward by multiples of 0.022 emu mol$^{-1}$ Oe$^{-1}$ for $H$ $\|$ $c$ and 0.0035 emu mol$^{-1}$ Oe$^{-1}$ for $H$ $\|$ $a$. The dashed lines guide the change of $T$$_N$ under magnetic fields. (d) Isothermal magnetization $M$($H$) curves for the magnetic field parallel to different crystallographic axes. The inset displays the weak magnetic transition at around 5.5 T. (e) Isothermal magnetization curves at different temperatures. (f) A temperature-magnetic
field phase diagram. The intensities of $\chi$$_{c}$ symbolize the magnetic phase boundary. The paramagnetic (PM), antiferromagnetic (AFM1, AFM2), and possible quantum paramagnetic (QPM) regions are determined by
susceptibility, specific heat, and isothermal magnetization measurements. The blue hollow and solid dots represent the critical fields for transitions observed during the decreasing and increasing magnetic field sweeps, respectively. The color represents the susceptibility shown in (c).}
\label{fig3}
\end{figure*}

Figure \ref{fig3}(a) shows the temperature dependence of the magnetic susceptibility with magnetic fields parallel to the crystallographic $a$, $b$, and $c$ axis. A small furcation exists between the zero-field-cooling (ZFC) and field-cooling (FC) susceptibility curves. A large difference among $\chi$$_{a}$, $\chi$$_{b}$, and $\chi$$_{c}$ suggests a strong magnetic anisotropy. Upon cooling, $\chi$$_{c}$ drops steeply at $T$$_{N2}$ = 16.3 and $T$$_{N1}$ = 14.3 K, indicating the developments of successive long-range AFM orders. No thermal hysteresis was observed in the magnetic susceptibility during heating and cooling cycles (see Fig. S1 in the supplemental material\cite{x1}). Figure \ref{fig3}(b) displays the inverse susceptibility. Above 100 K, the data can be fitted by the Curie-Weiss law\cite{Huang2021}, yielding a Weiss tepmerature $\theta$$_{CW}^{a}$ = $-$99.5 K and Curie constant $C$$^a$ = 6.16 emu K mol$^{-1}$ Oe$^{-1}$ for $H$ $\|$ $a$ axis; $\theta$$_{CW}^b$ = $-$25.2 K and $C$$^b$ = 5.98 emu K mol$^{-1}$ Oe$^{-1}$ for $H$ $\|$ $b$ axis; and $\theta$$_{CW}^c$ = $-$41.2 K and $C$$^c$ = 6.12 emu K mol$^{-1}$ Oe$^{-1}$ for $H$ $\|$ $c$ axis. The negative values of $\theta$$_{CW}$ suggest a dominant AFM-type interaction in this system. The effective magnetic moments of Co$^{2+}$ ions are calculated to be $\mu$$_{eff}^{a}$ = 4.96$\mu$$_B$, $\mu$$_{eff}^{b}$ = 4.89$\mu$$_B$, and $\mu$$_{eff}^c$ = 4.95$\mu$$_B$. All the effective moments are larger than 3.87$\mu$$_B$ of a free Co$^{2+}$ ion ($S$ = 3/2, $g$ = 2), suggesting that Co$^{2+}$ ions in the orthorhombic CaCo$_2$TeO$_6$  have a significant orbital moment. The evolutions of $\chi$$_{c}$ and $\chi$$_{a}$ at various magnetic fields are shown in Fig. \ref{fig3}(c). For $H$ $\|$ $c$, the anomalies at $T$$_{N1}$ and $T$$_{N2}$ shift toward the low-temperature side with increasing field and become flattened as the field up to 2.5 and 10 T, respectively, suggesting that magnetic field suppresses the long-range AFM orders. For $H$ $\|$ $a$, the magnetic order seems rather robust, and the magnetic transitions gradually merge into one as the field increases and persists at high fields.

Figure \ref{fig3}(d) displays the magnetization curves at selected temperatures. The magnetization process exhibits significant anisotropy at 2 K. The magnetization curves along the $H$ $\|$ $a$ and $H$ $\|$ $b$ axis are nearly linear, while successive jumps occur at $H$$_{c1}$ = 3.5 T, $H$$_{c2}$ = 5.3 T, and $H$$_{c3}$ = 9.4 T in the magnetization when magnetic fields along the $c$ axis. These jumps usually arise from the field-induced spin-flop transitions along the easy axis. In addition, magnetic hysteresis is evident around these transitions, consistent with the characteristics of a first-order transition. As the temperature increases, the hysteresis becomes weaker, and the spin-flop critical fields shift toward the lower fields, as shown in Fig. \ref{fig3}(e) and Fig. S2\cite{x1}.

Fig. \ref{fig3}(f) summarizes a phase diagram of the magnetic phase transitions. At the low-fields, the orthorhombic CaCo$_2$TeO$_6$ undergoes two AFM transitions from the paramagnetic (PM) state to the AFM1 ground state.
As the fields increase, the AFM1 state is tuned to an intermediate phase AFM2. The transition can be understood as a spin flop induced by the $H$$_{c1}$ field. An extra spin-flop transition can be
observed at $H$$_{c2}$ of around 5 T at the low-temperature region, possibly resulting from the reduction of spin fluctuations. When the field is enhanced to $H$$_{c3}$, the spins are further flopped to a disordered QPM state.

\begin{figure}[t]
\includegraphics[scale = 0.43]{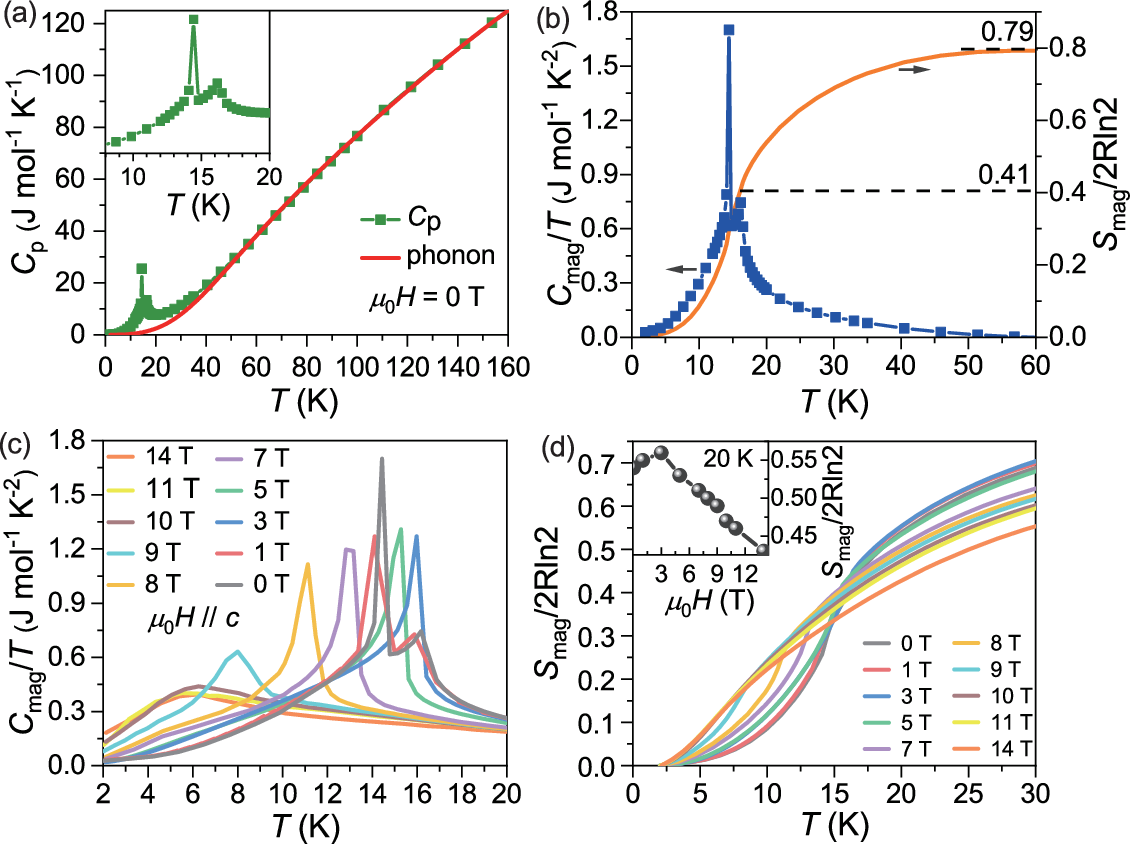}
\caption{ (a) Temperature dependence of the specific heat measured at zero field and a Debye fitting of the phonon mode. The inset shows a zoomed-in transition. (b) Magnetic specific heat and magnetic entropy as a function of temperature at zero field. Temperature dependence
of the (c) magnetic specific heat and (d) magnetic entropy in different magnetic fields along the $c$ axis. The inset shows magnetic entropy versus field along the $c$ axis at 20 K.}
\label{fig4}

\end{figure}

Figure \ref{fig4}(a) presents the temperature-dependent specific heat data in zero fields. The specific heat of the insulator includes the contributions of magnet part $C$$_{mag}$ and phonon part $C$$_{ph}$. Considering the coexistence of light atoms (such as O) and heavy atoms (such as Te), the phonon contribution can be estimated by the modified two-component Debye mode. The data are fitted well in the temperature range from 51 to 153 K using the following equation.
\begin{equation}
  C_{ph}= 9R\sum_{n=1}^{2}C_{n}(\frac{T}{\Theta_{Dn}})^3\int_{0}^{{\Theta_{Dn}}/T}\frac{x^4e^x}{(e^x-1)^2}\,dx
  \label{eq1}
 \end{equation}
 The fitting yields the number of heavy and light atoms of $C$$_1$ = 3.5 and $C$$_2$ = 6.5, consistent with the ten atoms in a formula unit of CaCo$_2$TeO$_6$. The Debye temperatures are $\Theta$$_{D1}$ = 287(1) K and $\Theta$$_{D2}$ = 882(3) K, corresponding to the heavy and light atoms, respectively. The fitted curve was then extrapolated to 2 K and subtracted from the measured specific heat data to obtain the magnetic contribution $C$$_{mag}$. The $C$$_{mag}$/$T$ data is shown in Fig. \ref{fig4}(b). Two transitions centered at $T$$_{N1}$ = 14.4 K and $T$$_{N2}$ = 16.2 K are observed, which suggests the establishment of two long-range magnetic orders. The integrated temperature-dependent magnetic entropy reaches 0.79$\times$2$R$ln2, suggesting an effective $J$$_{eff}$ of 1/2 for Co$^{2+}$ ions at low temperatures. The magnetic entropy released at the magnetic transition, $T$$_{N2}$, is only one-half of the full entropy (0.41$\times$2$R$ln2), suggesting the existence of large spin fluctuations. As shown in Fig. \ref{fig4}(c), the magnetic field applied along the $c$ axis can suppress the transitions gradually.
Specifically, the transitions at $T$$_{N1}$ and $T$$_{N2}$ can be suppressed by 3 and 10 T fields, respectively, which is consistent with the susceptibility data shown in Fig. \ref{fig3}(f). This suggests a field-induced spin-disordered state above 10 T. Figure \ref{fig4}(d) shows that the residual entropy is enhanced with increasing field, indicating a strengthening in the disorder degree of moments and frustration. The inset of Fig. \ref{fig4}(d) displays the field-dependence of magnetic entropy at 20 K. The tuning of magnetic entropy by field is nonmonotonic and reaches to a maximum at 3 T. This nonmonotonic behavior has also been observed in the well-known Kitaev spin liquid candidates as like  K$_2$Co$_2$TeO$_6$\cite{xu2024k2co2teo6}, Na$_2$Co$_2$TeO$_6$\cite{lin2021field}, and $\alpha$-RuCl$_3$\cite{Baek2017, Sears2017, Wolter2017}, which is related to the sequential closing and opening of an energy gap in magnetic excitations.

The magnetic couplings strongly depend on the bonding geometry based on the Goodenough-Kanamori rule\cite{Li2022, Goodenough1963}. The Co1-O2-Co1 and Co2-O7-Co2 bond angles are 98.51(16)$^\circ$ and 96.65(18)$^\circ$, respectively, smaller than 100$^\circ$, indicating dominated FM interactions between the adjacent Co1(Co2) ions. On the contrary, the adjacent Co1 and Co2 ions prefer AFM interactions due to the larger Co1-O5-Co2 (127.69(19)$^\circ$) and Co1-O3-Co2 (119.64(19)$^\circ$) bond angles. Besides, the 3.77 {\AA} of Co1-Co2 distance are significantly larger than 2.86 {\AA} of the Co1-Co1(Co2-Co2) along the $c$ axis. These structural features are consistent with the up-up-down-down magnetic structure, which reflect the underlying easy-axis magnetic anisotropy. The magnetic interactions along the $a$ axis are pure AFM type, consistent with the larger $\theta$$_{CW}$ than $\theta$$_{CW}^c$ derived from the magnetic susceptibilities.

In contrast to the recently reported rhombohedral phase of CaCo$_2$TeO$_6$\cite{PhysRevB.110.184411}, the crystals with orthorhombic structure exhibit distinct magnetic characteristics, including multiple magnetic order transitions and field-induced spin-flop transitions. This enhanced magnetic complexity likely stems from its reduced crystallographic symmetry (space group ${Pnma}$ vs. $R$$\bar{3}$). The magnetic interactions between CoO$_6$ octahedra are strongly influenced by trigonal distortions\cite{liu2020kitaev}, which can be characterized by the bond angle variance parameter\cite{Robinson1971}:
\begin{equation}
\sigma^{2}=\sum_{i=1}^{m}\dfrac{(\psi_{i}-\psi_{0})^{2}}{m-1}
\label{eq2}
\end{equation}
where $\psi$$_{0}$ is 90 deg representing an ideal octahedron with $O$$_h$ symmetry, $\psi$$_{i}$ is the individual bond angle within the CoO$_6$ octahedron, and $m$ denotes the number of $\psi$$_{i}$. Our calculations reveal significantly larger distortion parameters for the orthorhombic phase, with $\sigma$$^{2}$ = 70.7 deg$^{2}$ (Co1 site) and 95.7 deg$^{2}$ (Co2 site), compared to 36.02 deg$^{2}$ in the rhombohedral analog. Furthermore, the magnetic dimensionality exhibits a striking contrast between these polymorphs. The rhombohedral phase adopts a quasi-two-dimensional honeycomb architecture with well-separated Co layers, whereas the orthorhombic CaCo$_2$TeO$_6$ forms a three-dimensional network through reduced interlayer separation. The three-dimensional connection enhances interlayer exchange coupling. The pronounced octahedral distortion and emergent tridimensionality are expected to cooperatively suppress the relativistic Kitaev interactions while amplifying symmetric Heisenberg and off-diagonal exchange terms\cite{liu2020kitaev}.
This is indeed supported by the large entropy released near $T$$_N$, which accumulates to 0.79Rln2 up to 60 K, closing to the theoretical expectation of Rln2 for a Co$^{2+}$. In contrast, the entropy release in $\alpha$-RuCl$_3$ is dominated at a temperature well above $T$$_N$ that has been attributed to itinerant Majorana fermions associated with Kitaev spin liquid physics\cite{Widmann2019}. The emergence of the field-induced spin-disordered state in orthorhombic CaCo$_2$TeO$_6$ seems to originate from geometrical frustration as reported in BaCo$_2$(AsO$_4$)$_2$\cite{Thomas2023}. While the structure of the orthorhombic CaCo$_2$TeO$_6$ deviates from the pristine honeycomb structure, the competition of the exchange interactions in this complex structure calls for elucidation by inelastic neutron scattering.

\section{CONCLUSIONS}

We have systematically investigated the synthesis, magnetization, thermodynamics, and magnetic ground states of the orthorhombic CaCo$_2$TeO$_6$ single crystals. Magnetic susceptibility and specific heat show long-range AFM orders with transitions at $T$$_{N1}$ = 14.4 K and $T$$_{N2}$ = 16.2 K. The magnetic ground state is determined to be a canted $\uparrow$$\uparrow$$\downarrow$$\downarrow$ spin configuration characterized by a propagation vector $\bf{k}$$_1$ = (0, 0, 0) at 6 K. As an external magnetic field is applied along the easy $c$ axis, successive spin-flop transitions occur accompanied by the suppression of AFM orders and enhancement of residual magnetic entropy. The field-induced QPM state within this three-dimensional crystalline architecture is a fascinating phenomenon in quantum magnetism, as its emergence is related to the frustrated magnetic correlations.

\section{ACKNOWLEDGMENTS}

Work at SYSU was supported by the National Key Research and Development Program of China (Grant No. 2023YFA1406500), the National Natural Science Foundation of China (Grant Nos. 12304187 and 12425404), the Guangdong Basic and Applied Basic Research Funds (Grant No. 2024B1515020040), the open research fund of Songshan Lake Materials Laboratory (Grant No. 2023SLABFN30), the Guangzhou Basic and Applied Basic Research Funds (Grant Nos. 2024A04J4024 and 2024A04J6417), Guangdong Provincial Key Laboratory of Magnetoelectric Physics and Devices (Grant No. 2022B1212010008), and Research Center for Magnetoelectric Physics of Guangdong Province (Grant No. 2024B0303390001).

\bibliography{ref}
\end{document}